\documentstyle[12pt]{article}
\author{Serge Galam and Alain Mauger\\
Acoustique et Optique de la Mati\`{e}re Condens\'{e}e\footnotemark[1]\\
Tour 13 - Case 86, 4 place Jussieu, 75252 Paris Cedex 05, France\\[1ex].}
\title{A Quasi-exact Formula \\for \\Ising critical temperatures \\on \\Hypercubic Lattices  }
\addtocounter{footnote}{0}
\footnotetext{Laboratoire associ\'{e} au CNRS (URA n$^{\circ}$ 800) et
\`{a} l'Universit\'{e}
P. et M. Curie - Paris 6}
\addtocounter{footnote}{1}
\date{August 1996\\ 
(To appear in Physica A)}

\begin{document}
\maketitle
\begin{center}
{\em PA Classification Numbers:\/} 64.60 A, 64.60 C, 64.70 P\\
\end{center}

\begin{abstract}

We report a quasi-exact power law behavior for Ising critical temperatures 
on hypercubes. It reads $J/k_BT_c=K_0[(1-1/d)(q-1)]^a$
where $K_0=0.6260356$, $a=0.8633747$, $d$ is the space dimension, $q$ the coordination number 
($q= 2d$),
$J$ the coupling constant, $k_B$ the Boltzman constant and $T_c$ the critical temperature.
Absolute errors from available exact estimates ($d=2$ up to $d=7$)
are always less than $0.0005$.
Extension to other lattices is discussed.
 
\end{abstract}
\newpage

In ref. [1] we showed that both site and bond percolation thresholds
obey a universal power law,
\begin{equation}
p_c=p_0[(d-1)(q-1)]^{-a}d^{\ b}
\end{equation}
where $d$ is the space
dimension, $q$ the coordination number, $p_0$ and $a$ are constants.
For site dilution $b=0$ while $b=a$ for bond dilution. 

From a $Ln-Ln$ plot, all known percolation thresholds are found to align on 
two straight lines each one thus defining a universal class characterized by a set
of parameters $\{p_0; \ a\}$.
One class includes two-dimensional triangle, square
and honeycomb lattices, with
$\{p_0=0.8889; \ a=0.3601\}$ for
site dilution and by
$\{p_0=0.6558; \ a=0.6897\}$ for bond dilution. Two-dimensional Kagom\'{e}
and all lattices of cubic symmetry (for
$d\geq 3$) constitute the second class with 
$\{p_0=1.2868; \ a=0.6160\}$ and
$\{p_0=0.7541; \ a=0.9346\}$ for sites and bonds respectively.
At high dimensions a third class for hypercubes (sc and fcc) is introduced to
recover the infinite Cayley tree limit [1]. We noticed that some Ising critical temperatures were
found to obey Eq. (1) within the bond scheme ($b=a$).

Very recently Hackl and Morgenstern suggest a connection between on the one hand
bond percolation thresholds and, on the other one,a combination of Ising critical temperatures
and energies [2]. Agreement with available data are within few percents.

In this note we report a quasi-exact power law formula for Ising critical temperatures 
on hypercubes,
\begin{equation}
K_c=K_0[(1-\frac{1}{d})(q-1)]^{-a} \ ,
\end{equation}
where $K_c \equiv \frac{J}{k_BT_c}$, $J$ is the coupling constant, $k_B$ the Boltzman 
constant, $T_c$ the critical 
temperature, $q= 2d$, $K_0=0.6260356$ and $a=0.8633747$.
Absolute errors are always less than 0.0005. 
Application of Eq. (2) to other lattices is discussed.

Fig. (1) shows a $Ln-Ln$ plot of respectively $K_c^{-1}$ and $[(1-1/d)(q-1)]$ for all Ising
hypercubes from $d=2$ up to $d=7$. Data are exact estimates taken from refs. [3, 4]. 
All points (small squares in the Figure) align on a single straight line
which  demonstrates Ising critical temperatures ($K_c$) obey a power law of the form of Eq. (2).

Characteristics of that straight line are determined using  exact estimates $K_c^e$ from
$d=2$ up to $d=6$ not including $d=7$. We found $K_0=0.6260356$ and $a=0.8633747$.
Associated values of $K_c$ are reported in Table (1). The agreement with exact estimates 
is indeed remarkable. 
Defining $\Delta \equiv K_c-K_c^e$ where $K_c^e$ are exact estimates [3, 4] and $K_c$ are
from Eq. (2), absolute errors $|\Delta|$ are found to be always 
less than 0.0005.

Including $d=7$ yields $K_0=0.6269574$ and $a=0.8647815$ which still is excellent
but a little bit less accurate. The error $\Delta$ is then reduced for $d=7$ down 
to $\Delta =+0.00024$
but goes up to $\Delta =+0.00084$ for $d=2$ (see Table 1). This sensitivity to the case 
$d=7$ sheds light
on our conjecture of a crossover to a new universality class for percolation thresholds
above $d=6$ [1].
Moreover Eq. (2) with $a\neq 1$ does not have the $d \rightarrow \infty$ asymptotic Cayley 
tree limit $K_c = 1/q$ hinting to some crossover at high dimension.

In Fig. (2) Ising critical temperature exact estimates for non-hypercubic lattices 
$honeycomb$, $kagom\acute{e}$, $triangular$ at $d=2$, and  $diamond$, $bcc$ and $fcc$ at $d=3$
are included.
Data are seen not to sit exactly on the hypercubic straight line. Discrepancies are
between $\Delta=-0.031$ and $\Delta=0.0082$ (see Table 2). 
It hints some correction to $q=2d$ should be accounted for in Eq. (2). We found for instance
that rescaling the variable $(1-1/d)(q-1)$ by $(q/2d)^{0.20}$ reduces 
discrepancies for non-hypercubic systems. However this point needs more investigation.

To complete our presentation we have also determined 
$K_0=0.6247099$ and $a=0.8606241$ using only $d=2$ and $d=3$ data. Associated
errors $\Delta^*$ are shown in last column in Tables (1, 2). 

To conclude we have found a quasi-exact formula (Eq. 2) to yield
Ising critical temperatures on hypercubic lattices. At this stage it worthwhile 
to stress that a posteriori our results provide ground to technical efforts made to
determine critical temperature numerical
estimates up to more than six digits [3, 4]. Otherwise we would have 
conclude wrongly Eq. (2) is exact as seen from associated errors in Table (1).

More estimates at $d=8$ and $d=9$ are now needed to elucidate the high dimension crossover
to the  $d \rightarrow \infty$ asymptotic Cayley 
tree limit. Last but not least, it is interesting to note $[(1-\exp({-2K_c})]$ does not 
obey a power law in the variable $[(1-1/d)(q-1)]$.


\subsection*{Acknowledgments.}
We would like to thank Dietrich Stauffer for
stimulating discussions.

\vspace{3.0cm}
{\LARGE References}\\ \\
1. {\sf S. Galam and A. Mauger}, Phys.Rev.E
\underline {53},
2177 (1996)  \\
2. {\sf R. Hackl and I. Morgenstern }, Int.J.Mod.Phys.C\underline {7}, 609 (1996) \\
3. {\sf D. Stauffer and A. Aharony}, ``Introduction to Percolation Theory",
2nd Ed.,
Taylor and Francis, London (1994) \\
4. {\sf J. Adler}, ''Series versus simulations" in ''Annual Reviews of Computational
Physics, Vol.\underline {IV}, 241, ed. D. Stauffer (1996) \\


\newpage
{\LARGE Figure captions}\\ \\

Figure 1: $Ln-Ln$ plot of $K_c^{-1}$ versus  $\frac{(d-1)(q-1)}{d}$
for hypercubes from $d=2$ up to $d=7$. Small squares are exact estimates from [3, 4]. 
The straight line
is from Eq. (2).

Figure 2: $Ln-Ln$ plot of $K_c^{-1}$ versus $\frac{(d-1)(q-1)}{d}$
for hypercubes and other lattices from $d=2$ up to $d=7$. 
small squares are exact estimates from [3, 4]. The straight line
is from Eq. (2).

\newpage
\begin{table}

\label{tbl}
\begin{center}
\begin{tabular}{|l|l|r|r|r|r|r|}
\hline
Dimension &Lattice &$\,q$&${K_c^e}$&${K_c}$&$\Delta$ &$\Delta^*$\\ [5pt]
\hline
$d=2$ &Square & 4&  0.4406868& 0.44112972 & +0.00044&0\\ [5pt]
\hline

$d=3$ &sc& 6&  0.2216544& 0.22139040    &-0.00026&0\\ [5pt]
\hline
$d=4$ &sc& 8& 0.14970& 0.14956561  & -0.00013&+0.00023\\ [5pt]
\hline
$d=5$ &sc& 10& 0.1139291& 0.11386752& -0.00005&+0.00032\\ [5pt]
\hline
$d=6$ &sc& 12& 0.092295& 0.09228595  & +0.00014&+0.00051\\ [5pt]
\hline
$d=7$ &sc*& 14& 0.077706& 0.07809958& +0.00039&+0.00075\\ [5pt]
\hline
\end{tabular}
\end{center}
\caption{\sf Ising critical temperatures from this work $K_c$ compared to
``exact estimates"
$K_c^e$ taken from [3, 4]. $\Delta \equiv K_c-K_c^e$. $\ast$ means not
included to determine
$K_0=0.6260356$ and $a=0.8633747$. $\Delta^*$ corresponds to
$K_0=0.6247099$ and $a=0.8606241$ using only $d=2$ and $d=3$ data.
}
\end{table}

\begin{table}

\label{tbl}
\begin{center}
\begin{tabular}{|l|l|r|r|r|r|r|}
\hline
Dimension &Lattice &$\,q$&${K_c^e}$&${K_c}$&$\Delta$ &$\Delta^*$\\ [5pt]
\hline
$\,$ &Honeycomb* & 3&  0.6584788& 0.62603564  & -0.032&-0.034\\ [5pt]
$d=2$ &Kagom$\acute{e}$*&  4& 0.4665661 & 0.44112972  & -0.025&-0.026\\ [5pt]
$\,$ &Triangular* & 6&  0.2746531& 0.28381002   & +0.009&+0.009\\ [5pt]
\hline
$d=3$ &Diamond*& 4& 0.36982& 0.34411006 & -0.026&-0.026\\ [5pt]
$\,$ &bcc*& 8& 0.15740& 0.16557529   & +0.008&+0.008\\ [5pt]
$\,$ &fcc*& 12 & 0.10209& 0.11207783   & +0.010&+0.010\\ [5pt]
\hline
\end{tabular}
\end{center}
\caption{\sf Ising critical temperatures from this work $K_c$ compared to
``exact estimates"
$K_c^e$ taken from [3, 4]. $\Delta \equiv K_c-K_c^e$. 
$\ast$ means not
included to determine $K_0=0.6260356$ and $a=0.8633747$.
 $\Delta^*$ corresponds to
$K_0=0.6247099$ and $a=0.8606241$ using only hypercubic $d=2$ and $d=3$ data. 
}
\end{table}

\end{document}